\documentclass[prl, showpacs, superscriptaddress, twocolumn, floatfix]{revtex4}
\usepackage{color, graphicx, amssymb}  

\date{03/13/2009}
  
\begin{document}  

\title{Persistent currents in normal metal rings}

\author{Hendrik Bluhm}  
\altaffiliation[Present address: ]{Department of Physics, Harvard University, 
Cambridge, MA, USA}
\affiliation{Departments of Physics and Applied Physics, Stanford  
University, Stanford, CA 94305}  
\author{Nicholas C. Koshnick}  
\affiliation{Departments of Physics and Applied Physics, Stanford  
University, Stanford, CA 94305}  
\author{Julie A. Bert}  
\affiliation{Departments of Physics and Applied Physics, Stanford  
University, Stanford, CA 94305}  
\author{Martin E. Huber}  
\affiliation{Departments of Physics and Electrical Engineering,
University of Colorado Denver, Denver, CO 80217}  
\author{Kathryn A. Moler}
\email{kmoler@stanford.edu}    
\affiliation{Departments of Physics and Applied Physics, Stanford  
University, Stanford, CA 94305}

\begin{abstract}
The authors have measured the magnetic response of 33 individual cold
mesoscopic gold rings, one ring at a time.  The response of some
sufficiently small rings has a component that is periodic in the flux
through the ring and is attributed to a persistent current. Its period
is close to $h/e$, and its sign and amplitude vary between rings.  The
amplitude distribution agrees well with predictions for the typical
$h/e$ current in diffusive rings. The temperature dependence of the
amplitude, measured for four rings, is also consistent with
theory. These results disagree with previous measurements of three
individual metal rings that showed a much larger periodic response
than expected.  The use of a scanning SQUID microscope enabled in situ
measurements of the sensor background.  A paramagnetic linear
susceptibility and a poorly understood anomaly around zero field are
attributed to defect spins.
\end{abstract}

\maketitle  
When a conducting ring is threaded by a magnetic flux $\Phi_a$, the
associated vector potential imposes a phase gradient on the electronic
wave functions, $\psi$, that can be transformed into a phase factor in
the boundary conditions: $\psi(L) = e^{i 2 \pi \Phi_a/\phi_0}
\psi(0)$, where $L$ is the circumference of the ring and $\phi_0
\equiv h/e$ the flux quantum \cite{ByersN:Theccq}.  The $h/e$
periodicity of this phase factor is reflected in all properties of the
system. Here, we focus on the persistent current $I$ circulating the
ring, which is the first derivative of the free energy with respect to
$\Phi_a$, and thus a fundamental thermodynamical quantity.  For a
perfect 1D ring without disorder populated by noninteracting
electrons, it is relatively straightforward to show that $I$ will be
of order $e v_F/L$ \cite{CheungHF:Percso}, the current carried by a
single electron circulating the ring at the Fermi velocity $v_F$.
Perhaps somewhat surprisingly, persistent currents are not destroyed
by elastic scattering \cite{ButtikerM:Josbsn}.  In the diffusive limit,
i.e. for a mean free path $l_e < L$, $I \sim e/\tau_D$ is set by the
diffusive round trip time $\tau_D = L^2/D$, where $D = v_F l_e/3$ is
the diffusion constant \cite{CheungHF:Percmr, RiedelEK:Mespcs}.
Thermal averaging leads to a strong suppression of the persistent
current at temperatures above the correlation energy $E_c \equiv \hbar
\pi^2 D/L^2 \propto \hbar/\tau_D$.

Like many mesoscopic effects in disordered systems, the persistent
current depends on the particular realization of disorder and thus
varies between nominally identical samples.  In metal rings, the
dependence on disorder and $\cos (k_F L)$, which is random in
practice, leads to a zero ensemble average $\langle I_{h/e}\rangle$ of
the first, i.e. $h/e$-periodic, harmonic.
The magnitude of the fluctuations from sample to sample is given by the typical
value  \cite{RiedelEK:Mespcs}
\begin{equation}\label{eq:theory}
\langle I_{h/e}^2 \rangle^{1/2} = \frac{E_c}{\phi_0} e^{-k_B T/E_c}.
\end{equation}
We have not included a factor 2 for spin because our Au rings are 
in the strong spin-orbit scattering limit
\cite{MeirY:Uniess, Entin-WohlmanO:Effssm}.
An additional contribution that survives averaging over disorder 
but oscillates with $k_F L$ \cite{BaryS:avge} is predicted to have a magnitude
$\overline{I_{h/e}} = (12/\pi^2) \sqrt{M L/l_e} (E_c/\phi_0) e^{-2L/l_e}$,
where $M$ is the number of channels 
\cite{CheungHF:Percmr}.
Because of the exponential dependence on $L/l_e$, it is usually 
negligible compared to Eq. \ref{eq:theory} for metallic rings.
Higher harmonics are generally smaller because they are more sensitive
to disorder and thermal averaging.  However, due to interactions
\cite{AmbegaokarV:Cohapc, SchmidA:Percmr, SchechterM:Magrdm, Bary-SorokerH:Effpbm} 
and differences between the canonical and grand canonical ensemble 
\cite{AltshulerBL:Perdbc, VonOppenF:Avepcm, SchmidA:Percmr}, 
$\langle I_{h/2e}\rangle$ is expected to be nonzero.

There are very few experimental results on persistent
currents, and most measured the total response of an ensemble of
rings \cite{LevyLP:Magmcr, ReuletB:DynriA, DeblockR:aceam,
DeblockR:Diaorm}. The experiments  to date are all based on magnetic
detection and are considered challenging as they require a very high
sensitivity. The measurements of large ensembles are dominated by
$\langle I_{h/2e}\rangle$, whose contribution to the total current of
$N$ rings scales with $N$, whereas the $h/e$ periodic current scales
as $\sqrt{N}$ because of its random sign.
The measured values of $\langle I_{h/2e}\rangle$ are generally a factor 
of a few larger than  most theoretical predictions. A
plausible reconciliation was proposed 
recently for metallic rings \cite{Bary-SorokerH:Effpbm}.

Here, we address $\langle I_{h/e}^2 \rangle$ in diffusive rings by measuring
one ring at a time. The $h/e$ component has been measured in good
agreement with theory \cite{CheungHF:Percso} in a single ballistic
ring \cite{MaillyD:Expopc} and an ensemble of 16 nearly ballistic
rings \cite{RabaudW:Percmc} in semiconductor samples.  Measurements of
three diffusive metal rings \cite{ChandrasekharV:Magrsi} on the other
hand showed periodic signals that were 10--200 times larger than
predicted \cite{RiedelEK:Mespcs}.  Later results on the total current
of 30 diffusive rings \cite{JariwalaEMQ:Diapcd} showed a better
agreement with theory \cite{RiedelEK:Mespcs}, but did not allow one to
distinguish between the typical and average current, which would
require individual measurements of several rings or groups of
rings. Thus, there is an unresolved contradiction between experiment
and theory for the typical $h/e$ current, the investigation of which
is a major open challenge in mesoscopic physics.

We report measurements of the individual magnetic responses of
33 diffusive Au rings. 
The use of a scanning SQUID 
technique allowed us to measure many different rings, one by one, with
in situ background measurements \cite{HuberM:Susc,KoshnickN:Flusm}.
The response of some of the rings contains an $h/e$ periodic component
whose amplitude distribution -- including rings without a detectable
periodic signal -- is in good agreement with predictions for
$\langle I_{h/e}^2 \rangle^{1/2}$.  Additional features in the total
nonlinear response most likely reflect a nonequilibrium response of
impurity spins.  Different frequency and geometry dependencies allow
the distinction between those two components, and support the
interpretation of the periodic part as persistent currents.
Due to the necessity to subtract a mean background from our data and the small
number of rings, we are unable to extract any ensemble average from
our results.

\begin{figure}
\includegraphics{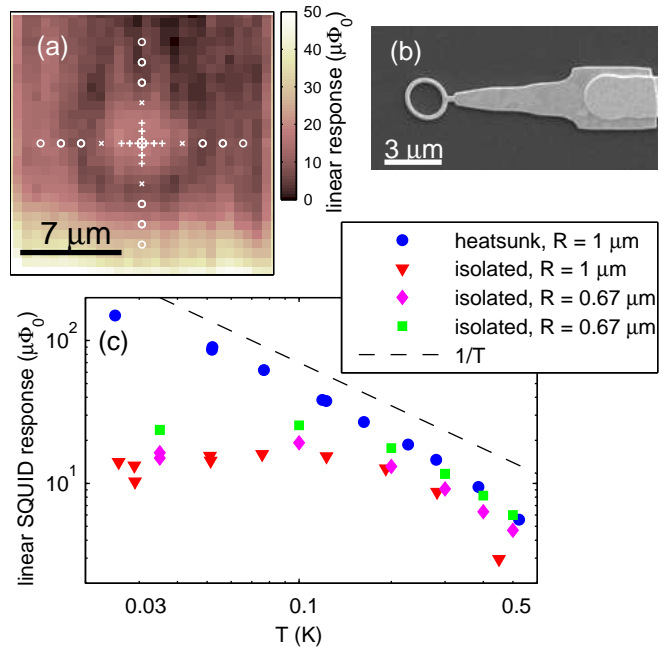}  
\caption{\label{fig:scan}(color online).
(a) Susceptibility scan of an isolated ring used to locate the ring and 
to determine the indicated measurement positions. Background measurements 
at positions ``o'' are subtracted from the data taken at positions ``+'' 
to obtain the ring response.
(b) Scanning electron micrograph of a heatsunk ring. 
(c) Temperature dependence of the linear response of one heatsunk and three 
isolated rings. The data in (a) and (c) reflect the total amplitude
of the linear response to a sinusoidal excitation of $\pm$45 G for (a) and 
the 0.67 $\mu$m rings in (c), and $\pm$35 G for the 1 $\mu$m rings.
}
\end{figure}

Our samples were fabricated using standard e-beam and optical
liftoff lithography and were e-beam evaporated from a 99.9999 \% pure Au 
source onto a Si substrate with a native oxide.
The 140 nm thick rings were deposited
at a relatively high rate of 1.2 nm/s in order to achieve a large
$l_e$. The rings have an annulus width of 350 nm,
and radii $R$ from 0.57 to 1 $\mu$m.  From resistance
measurements of wires fabricated together with the rings, 
we obtain $D$ = 0.09 m$^2$/s, $l_e$ = 190 nm.
Weak localization measurements yield a dephasing length
$L_\phi$ = 16 $\mu$m  at $T$ = 300 mK, so that $ L_\phi \approx 4 L$
for our most important $R$ = 0.67 $\mu$m rings.
Some rings were connected to large metallic banks 
[See Fig. \ref{fig:scan}(b)] to absorb the inductively coupled 
heat load from the sensor SQUID.

The experiment was carried out using a dilution-refrigerator based
scanning SQUID microscope \cite{BjornssonPG:Scasqi}. 
Our sensors \cite{HuberM:Susc} have an integrated field coil of 
13 $\mu$m mean diameter, which is used to apply a field to the sample.
The sample response is coupled into the SQUID via a 4.6 $\mu$m diameter 
pickup loop. 
A second, counter-wound pair of coils 
cancels the response to the applied field 
to within one part in $10^4$ \cite{HuberM:Susc}.
The sensor response to a current $I$ in a ring is $\Phi_{SQUID} = M I$, where 
$M$ is the pickup-loop--ring inductance. 
Independent estimates based on previous experiments 
\cite{KoshnickN:Flusm,BluhmH:Magrms}
and modeling give $M = R^2 \cdot 0.3 \Phi_0/\mu$m$^2$mA, where 
$\Phi_0 \equiv h/2e$ is the superconducting flux quantum.
Using the measured $D$,  Eq. \ref{eq:theory} thus predicts a typical
$h/e$ response from persistent currents of 
$M \langle I_{h/e}^2 \rangle^{1/2} = 0.17\; \mu\Phi_0 \cdot e^{-k_B T/E_c}$,
where $\Phi_0 \equiv h/2e$ is the superconducting flux quantum.
We neglect the contribution from $\overline{I_{h/e}}$, which is 
0.3 $E_c/\phi_0$ for our smallest rings and much less for larger rings.

After coarse alignment by imaging a current-carrying
meander wire on the sample, accurately locating a ring is
facilitated by a paramagnetic susceptibility of our metal structures
that appears in scans of the linear response to an applied field
[Fig.  \ref{fig:scan}(a)].  
To measure the complete nonlinear response, we digitized the SQUID
signal at a sample rate of 333 kHz and averaged it over many sweeps of 
the current through the field coil, which was varied sinusoidally 
over the full field range at typically 111 Hz.  
This raw signal of a few m$\Phi_0$ is
dominated by nonlinearities in the sensor background and a small phase
shift between the fluxes applied to the two pickup loop--field coil
pairs. To extract the response of a ring, we measured at the positions
indicated in Fig. \ref{fig:scan}(a), and subtracted datasets taken far
from the ring (o) from those near the ring (+). The total 
averaging time for each ring was on the order of 12 hours. 
The reduced coupling 
to the SQUID at intermediate positions was accounted for through a
smaller prefactor.  The symmetric measurement positions eliminate
linear variations of the sensor background, which in some cases are
larger and more irregular than the final signal.  The reliability of
the final result can be assessed by checking if its features (typically
characterized by higher harmonics of the sensor response) show a
spatial dependence similar to that of the ring--pickup-loop coupling.
This check allowed us to identify and discard questionable datasets 
with very irregular features of ~1 $\mu\Phi_0$ obtained  in some sample 
regions.
 
The response of our rings is dominated by a paramagnetic linear
component of up to $\approx$150 $\mu\Phi_0$ at a field of 45 G
\cite{BluhmH:Susc}.  Its temperature dependence is
shown in Fig. \ref{fig:scan}(c).
 The linear response of heatsunk rings and heatsinks
\cite{BluhmH:Susc} (not shown) varies approximately as $1/T$. Thus, it
is likely due to spins.  Its magnitude corresponds to a density of $4
\cdot 10^{17}$ spins/m$^2$ (assuming spin 1/2).  If these spins were
identical to the metallic magnetic impurities that were shown
to cause excess dephasing \cite{PierreF:Depemm}, one would expect a
much larger spin flip dephasing rate than the upper bound obtained
from our $\tau_\phi$ measurements.

The linear response of isolated rings varies little below $\approx$ 150 mK.
This indication of a saturating electron temperature agrees
with estimates of the heating effect 
of the $~$10 $\mu$A, 10 GHz Josephson current in the SQUID pickup loop
\cite{aux}.
The different behavior of 
heatsunk  and isolated rings shows that the linear susceptibility
reflects the electron rather than phonon temperature.

\begin{figure}
\includegraphics{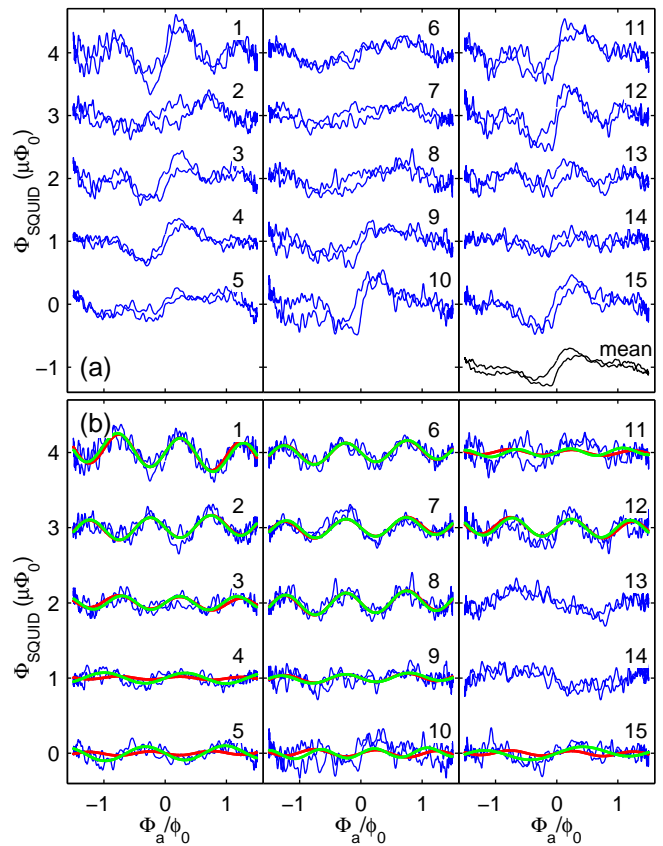}
\caption{\label{fig:ensemble}(color online).
(a) Response of 15 nominally identical rings with $R$ = 0.67 $\mu$m after 
subtracting the in- and out-of-phase component of the linear response.
The curve labeled ``mean'' is the average of datasets 1-15.
(b) Results of subtracting this mean from datasets 1-15 in (a).
The smooth lines are sinusoidal fits (including a linear background term)
with fixed (red/dark) and fitted period (green/light). 
Datasets 13 and 14 were excluded from the analysis because of their stronger  
zero field anomaly.
The rms amplitude estimated from the fixed and variable period fits
corresponds to a current of 0.8 and 0.9 nA, respectively, in agreement with 
the expected value of $\langle I_{h/e}^2 \rangle^{1/2}$ 
from  Eq. \ref{eq:theory}.
}
\end{figure}

We now focus on the much smaller 
nonlinear response, obtained after eliminating the linear response 
(including a component that is out-of-phase with the sinusoidal applied field) 
by subtracting a fitted ellipse.
This linear component varied by up to a factor 2 between
nominally identical rings.
Fig. \ref{fig:ensemble}(a) shows data from fifteen isolated 
rings with $R$ = 0.67 $\mu$m.
While these raw data are not periodic in $\Phi_a$, most of them 
can be described as the sum of a periodic component and a 
step-like shape near $\Phi_a = 0$.
This unexpected, 
poorly understood anomaly appeared in nearly all rings, and was 
most pronounced in heatsunk rings \cite{aux}. 
Its frequency dependence suggests that it is 
due to nonequilibrium effects in the spin response, but it might 
mask a persistent-current-like effect \cite{aux}.

Since one might expect the same spin signal from each ring, whereas
persistent currents should fluctuate around a zero mean, we subtracted
the average of all fifteen datasets from each individual curve.  The
results [Fig. \ref{fig:ensemble}(b)] show oscillations that can be
fitted with a sine curve of the expected period for most rings.
Datasets 4, 5 and 15 give better fits with a 30 \% larger period,
which corresponds to an effective radius close to the ring's inner
radius. This variation of the period may reflect an imperfect
background elimination, but could also be a mesoscopic fluctuation of
the effective ring radius.
The seemingly much larger period of datasets 13 and 14 appears to be
due to a different magnitude of the zero field anomaly.  From the sine
curve fits to 13 datasets, we obtain an estimate for $M \langle
I_{h/e}^2 \rangle^{1/2}$ of 0.11 $\mu\Phi_0$ if fixing the period at
the value expected for the mean radius of the rings, or 0.12
$\mu\Phi_0$ if treating it as a free parameter.  This value agrees
with the theoretical value of 0.12 $\mu\Phi_0$ from Eq.
\ref{eq:theory} for $T$ = 150 mK, which corresponds to $\langle
I_{h/e}^2 \rangle^{1/2} = 0.9$ nA for $R$ = 0.67 $\mu$m.

We checked the reproducibility of the response over several weeks
without warming up the sample for seven rings, and found good
consistency in five cases.  Reducing the field sweep range from 45 to
35 and 25 G or varying the frequency between 13 and 333 Hz changed the
step feature, but had little effect on the oscillatory component in
the difference between the responses of two rings \cite{aux}.

Out of five measurements of rings with $R$ = 0.57 $\mu$m \cite{aux}, four gave
similar results after subtracting their mean response as the $R$ =
0.67 $\mu$m rings.  The rms value of the fitted sine amplitudes was
0.06 and 0.07 $\mu\Phi_0$ for variable and fixed period, respectively.
A fifth ring was excluded from this analysis because it had a
significantly larger zero field anomaly.  Data from additional three
rings were rejected because of a large variation of the sensor
background that was not connected with the rings.

We also measured eight isolated rings with $R$ = 1 $\mu$m, which are
expected to give a smaller signal because of their smaller $E_c$ of
170 mK and stronger heating from the SQUID \cite{aux}.  Since the
magnitude of the zero-field anomaly varies significantly for these
rings, the mean subtraction procedure cannot fully remove it.  One of
these rings shows a sinusoidal signal with a period of 1 to 1.15
$\phi_0$ and an amplitude of up to 0.1 $\mu\Phi_0$, but poor
reproducibility.  Fitting sine curves, regardless of the absence of
clear oscillations for the other seven rings, gives $M \langle
I_{h/e}^2 \rangle^{1/2}$ = 0.03 $\mu\Phi_0$.  None of those rings show
a signal at a period similar to those in Fig. \ref{fig:ensemble}. This
dependence of the signal on the ring size supports the interpretation
as persistent current, as opposed to an artifact of the spin response.

\begin{figure}
\includegraphics{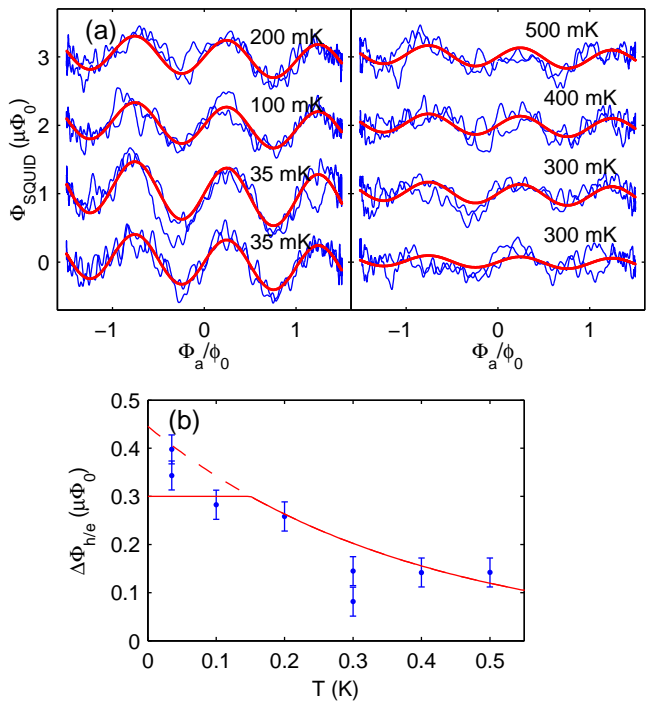}  
\caption{\label{fig:tempdep}(color online).
(a) Difference between the nonlinear responses of  two rings with
a large oscillatory component (curves 1 and 2 in
Fig. \ref{fig:ensemble}) at $T$ = 0.035 K to 0.5 K.
(b) Temperature dependence of the amplitude of the sinusoidal fits 
in panel (a). The exponential curve is a fit to 
$\exp(-\mathrm{min}(T, 0.15 K)/0.38 K)$, taking the
the saturation of the electron temperature into account.
The error bars were obtained by analyzing the $x$ and $y$ scan across the 
rings  [cf. Fig. \ref{fig:scan}(a)] separately and averaging the 
difference square of the respective results over all eight data points.
}
\end{figure}

The data discussed so far was taken at base temperature.
We have measured the temperature dependence of the responses of
four 0.67 $\mu$m rings with large oscillatory signals of opposite sign.
Taking the difference between their nonlinear responses, 
which eliminates any common background signal, 
leads to predominantly sinusoidal curves at most temperatures,
as shown in Fig. \ref{fig:tempdep}(a).
The period appears to be $T$-independent, and  amplitudes from 
fits with a fixed period are consistent with an $e^{-k_B T/E_c}$
dependence [Fig. \ref{fig:tempdep}(b)] with $E_c/k_B$ = 380 mK, 
as obtained from the measured $D$.

In this experiment, the $h/e$ persistent current in diffusive 
rings is in good agreement with theory within the temperature range covered,
providing long-overdue experimental input to the questions raised 
by an earlier experiment \cite{ChandrasekharV:Magrsi}.

\acknowledgments{This work was supported by NSF Grants No.
DMR-0507931, DMR-0216470, ECS-0210877 and PHY-0425897 and by the 
Packard Foundation. Work was performed in part at the Stanford 
Nanofabrication Facility, supported by NSF Grant No. ECS-9731293.
We would like to thank Yoseph 
Imry, Moshe Schechter and Hamutal Bary-Soroker for useful discussions.}

\bibliography{pc_bibdata,susc_bibdata}
\end{document}